\documentclass[aps,twocolumn,nofootinbib,floatfix]{revtex4-2}
\usepackage{graphicx}
\usepackage{epsfig}
\usepackage{epstopdf}
\usepackage{amsfonts}
\usepackage{amssymb}
\usepackage{amsbsy}
\usepackage{amsmath}
\usepackage{mathrsfs}
\usepackage{latexsym}
\usepackage{natbib}
\usepackage{bm}
\usepackage{soul}
\usepackage{color}
\usepackage{braket}
\usepackage{slashed}
\usepackage{pgfplots}
\usepackage[normalem]{ulem}

\usepackage{tikz}
\usetikzlibrary{shapes,arrows,shadows}
\usepackage{hyperref,url}
\hypersetup{
	colorlinks=true,       
	linkcolor=blue,          
	citecolor=blue,        
	filecolor=blue,      
	urlcolor=blue           
}

\begin{document}


\date{\today}

\title{Gyromagnetic ratio and Pauli form factor in anisotropic QED$_{2 + 1}$ at one-loop}

\author{Susobhan Mandal}
\email{sm12ms085@gmail.com}
\affiliation{Department of Physics, Indian Institute of Technology Bombay, Mumbai 400076, India}

\author{Subrata Mandal}
\email{subrata.mandal@itp3.uni-stuttgart.de}
\affiliation{Institute for Theoretical Physics III, University of Stuttgart, 70550 Stuttgart, Germany}

\begin{abstract}
The gyromagnetic ratio and Pauli form factor are important quantities that characterize the electric and magnetic moment distribution of a particle. The experimentally measured value for the gyromagnetic ratio or the spin g-factor of an electron in a vacuum agrees remarkably well with theoretical predictions, making it one of the biggest successes of quantum field theory. However, these factors may get modified compared to their vacuum counterparts due to the interaction present in the system. Motivating from that, in this article, we investigate the effects of interactions within the framework of $2+1$ dimensional Proca quantum electrodynamics. We demonstrate how the g-factor and Pauli form factors change with the electron's Fermi velocity and the mass of the vector fields within Proca quantum electrodynamics.
\end{abstract}

\maketitle

\section{Introduction}

In this article we explore the role of interactions and space-time anisotropy on the intrinsic g-factor and Pauli form factor in two-dimensional electronic systems. 
The anisotropy of space-time generally leads to a violation of Lorentz symmetry. In the field theories of two-dimensional electronic systems this anisotropy is essentially captured by the Fermi-velocity of Dirac fermions \cite{RevModPhys.81.109}. A similar situation arises when relativistic field theories are coupled to a thermal bath, where the system naturally selects a preferred time-like direction \cite{Das:1997gg, Kapusta:2006pm, Schmitt:2014eka}. As a result, in such systems, both the interaction and anisotropy play a crucial role in determining the electromagnetic form factors of Dirac fermions.
%

The g-factor, or gyromagnetic ratio, is a dimensionless quantity that characterizes the magnetic moment and angular momentum of particles such as electrons. On the other hand, the form factor describes the electric and magnetic moment distribution within a particle. These factors are crucial in particle physics because they provide insight into fundamental interactions and the structure of particles. The g-factor, in particular, has been precisely measured for the electron and compared with theoretical predictions \cite{PhysRev.73.416,
PhysRevLett.59.26,PhysRevLett.100.120801,PhysRevLett.109.111807}. The agreement has remained one of the greatest triumphs of the quantum field theory. The description of electron-electron interactions in 2D materials has previously been captured in pseudo-quantum electrodynamics (PQED) \cite{MARINO1993551,PhysRevB.95.245138}. Here, we use a similar anisotropic quantum electrodynamics (QED) framework that does not treat space and time equally to calculate the spin gyromagnetic factor and the Pauli form factor for two-dimensional electronic systems with short-range interactions. Specifically, we consider the static charges feel $K_{0}(mr)$ potential at the tree-level. Another significant difference in our approach is the incorporation of massive photons-described by the Proca fields \cite{proca1936theorie,PhysRev.76.66} which mediate interactions in $(2+1)$D and constrain the massive vector field to remain confined to the plane, in contrast to PQED which exploits a projected electromagnetic field in describing the properties of $(3+1)$D photons. Our results highlight the importance of interactions and velocity of Dirac particles in shaping the form factors. 

%

This work is organized as follows: in Sec. II, we present a general introduction to form factors in QED$_{ 3+1}$ theory followed by Sec. III, where we introduce the anisotropic QED$_{ 2+1}$ with massive vector fields. In Sec. IV, we derive the form factors for our model within a one-loop scheme before summarizing our findings in Sec. V.

\section{Form factors in QED$_{3 + 1}$}\label{section 1}
\begin{figure}[b]
\includegraphics[scale=0.5]{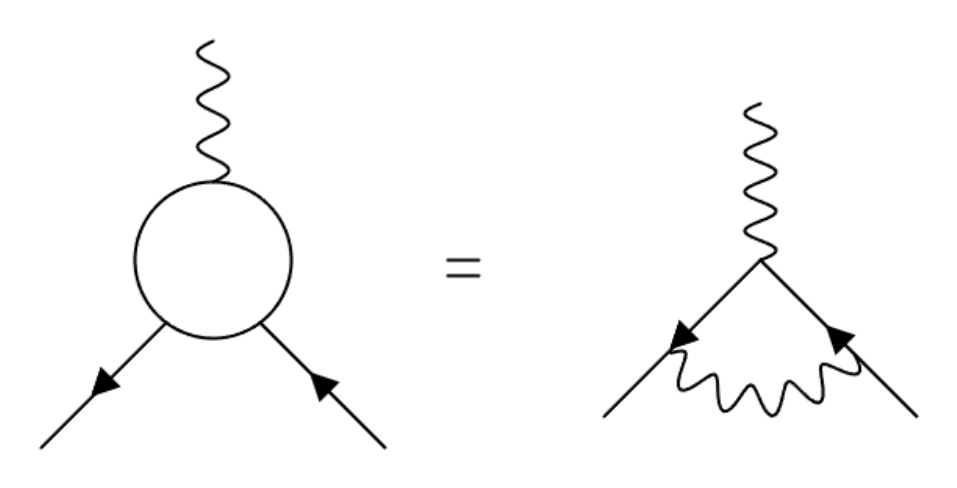}
\caption{Feynman diagram for electron-electron scattering}
\label{FIG. 1}
\end{figure}

In this section, we will briefly review the idea of gyromagnetic ratio and form factor in the context of QED. To begin with, let us consider a scenario where an electron is elastically scattered off a nucleus, with momenta transfer of $|\vec{q} |$ to the nucleus. Now, in the case of $|\vec{q} |$ being large, one can expect that the scattering process relies both on the total electric charge $Ze$ of the nucleus and the spatial charge distribution within the nucleus. In the first Born approximation,  the scattering amplitude can be shown to have the following form: 

\begin{equation}\label{eq. 2.1}
f(\vec{q}) = \frac{em_{e}}{2\pi}A^{0}(\vec{q}),
\end{equation} 
where
\begin{equation}\label{eq. 2.2}
A^{0}(\vec{q}) = \frac{\rho(\vec{q})}{\vec{q}^{2}} = \frac{1}{\vec{q}^{2}}\int d^{3}x \rho(\vec{x})
e^{-i\vec{q}.\vec{x}},
\end{equation}
with $\rho(\vec{x})$ denoting the charge distribution of the nucleus, and $e$, $m_e$ presenting the charge and mass of electron, respectively. Rewriting 
\begin{equation}\label{eq. 2.3}
A^{0}(\vec{q}) = \frac{Ze}{\vec{q}^{2}}F(\vec{q}^{2}),
\end{equation}
with 
\begin{equation}
\ F(\vec{q}^{2}) = \frac{1}{Ze}\int d^{3}x \rho(\vec{x})e^{-i\vec{q}.\vec{x}},
\end{equation}
the scattering amplitude can be expressed as
\begin{equation}\label{eq. 2.4}
f(\vec{q}^{2}) = \frac{Ze^{2}m_{e}}{2\pi\vec{q}^{2}}F(\vec{q}^{2}).
\end{equation}
Here, $F(\vec{q}^{2})$ is widely known as the nuclear form factor, which captures the deviation of the nuclear charge distribution from a point-like structure. On the other hand, in the limit of small momentum transfer, i.e., $q \rightarrow 0$, the form factor remains close to unity. Therefore, form factors are considered to be important quantities for unveiling the nuclear structure in the limit of large momentum transfer. In principle, one can understand the charge distribution of any particle from the form factors that are experimentally determined from scattering amplitudes. 
This sets up the premise for our work. Here, we discuss the form factors of Dirac particles in two dimensions. Fundamentally, these form factors can be derived from the electron-electron scattering. This scattering amplitude can be obtained from the Feynman diagram, shown in FIG.\ref{FIG. 1}, and given by 
\\
\begin{equation}\label{eq. 2.5}
i\mathcal{M} = - i e \times \bar{u}(p')\Gamma^{\mu}(p',p)u(p),
\end{equation}

where the electron-photon-electron vertex is $(- ie\Gamma^{\mu})$. Clearly, the vertex function, $(- ie\Gamma^{\mu})$ is a function of the initial and final electron momenta, $p$ and $p'$, as well as the momentum transfer $q=p'-p$. This function provides a set of form factors that are labeled by Lorentz and Dirac indices $\mu, \alpha, \beta$. With on-shell electron momenta $p$ and $p'$ and the photon momentum approaching zero ($q \rightarrow 0$), the form factor reduces to $1$, leading to our familiar form
\begin{equation}\label{eq. 2.6}
\Gamma_{\alpha\beta}^{\mu} \rightarrow \gamma_{\alpha\beta}^{\mu}.
\end{equation}
A more interesting situation arises when momentum transfer is non-zero. 
In that case, to determine the form factors, one exploits the Lorentz symmetry. It is evident that the $\Gamma^{\mu}_{\alpha\beta}(p,p')$ is a vector-valued function, and therefore, it must have form \cite{peskin2018introduction}
\begin{equation}\label{eq. 2.26}
\Gamma^{\mu}(p', p) = F_{1}(q^{2})\times\gamma^{\mu} + F_{2}(q^{2})\times\frac{i\sigma^{\mu\nu}
q_{\nu}}{2M},
\end{equation}
where $F_{1}(q^{2})$ and $F_{2}(q^{2})$ are known as electric and magnetic form factors.

\section{Anisotropic QED$_{2 + 1}$}\label{section 2}
The anisotropic QED$_{2 + 1}$ can be captured by the following action
\begin{equation}\label{eq. 3.1}
\begin{split}
S & = \int d^{3}x \ \Big[ - \frac{1}{4}F_{\mu\nu}F^{\mu\nu} + \bar{\psi}(i\gamma^{0}\partial_{0}
 + iv_{F}\gamma^{i}\partial_{i} - m)\psi\\
 & - e\bar{\psi}\left(\gamma^{0}A_{0} + v_{F}\gamma^{i}A_{i}\right)\psi + \frac{1}{2}m_{\gamma}^{2}
 A_{\mu}A^{\mu} - \frac{\lambda}{2}(\partial_{\mu}A^{\mu})^{2}\Big],
\end{split}
\end{equation}
where $\psi$ and $\bar{\psi}= {\psi}^{\dagger} {\gamma}_0$ are four-component Dirac spinors,  $v_{F}$ is the Fermi-velocity of electrons confined in two-dimensional plane. $F_{\mu\nu}
= \partial_{\mu}A_{\nu} - \partial_{\nu}A_{\mu}$ is the electromagnetic field-strength tensor, and $m_{\gamma}$ is mass of the gauge boson. $\gamma^{\mu}$s are the Dirac gamma matrices satisfying the Clifford algebra $\{\gamma^{\mu},
\gamma^{\nu}\} = 2\eta^{\mu\nu}\mathbf{1}_{3\times 3}$ with $\eta_{\mu\nu} = \text{diag}(1, -1, -1)$ being the Minkowski metric in the $2 + 1$ dimension. Note that the last term of the above action (\ref{eq. 3.1}) is the gauge fixing term.

Introducing the scaled gamma matrices in the following manner
\begin{equation}\label{eq. 3.2}
\Gamma^{0} = \gamma^{0}, \ \Gamma^{i} = v_{F}\gamma^{i},
\end{equation}
we may express the action (\ref{eq. 3.1}) as
\begin{equation}\label{eq. 3.3}
\begin{split}
S & = \int d^{3}x \ \Big[ - \frac{1}{4}F_{\mu\nu}F^{\mu\nu} + \bar{\psi}(i\Gamma^{\mu}
\partial_{\mu} - m)\psi - e\bar{\psi}\Gamma^{\mu}A_{\mu}\psi\\
 & - \frac{\lambda}{2}(\partial_{\mu}A^{\mu})^{2} + \frac{1}{2}m_{\gamma}^{2}A_{\mu}A^{\mu}\Big],
\end{split}
\end{equation}
where the $\Gamma^{\mu}$ matrices satisfy the following modified Clifford algebra
\begin{equation}\label{eq. 3.4}
\{\Gamma^{\mu}, \Gamma^{\nu}\} = 2M^{\mu\nu}\mathbf{1}_{2\times 2},
\end{equation}
where $M^{\mu\nu} = \text{diag}(1, - v_{F}^{2}, - v_{F}^{2})$. We use this notation for computing the form factors in the next section. In order to determine the electromagnetic 
form factors, we will use the following identities 
\begin{equation}\label{eq. 3.5}
\begin{split}
\Gamma^{\nu}\slashed{a}\Gamma_{\nu} & = \slashed{a}(1 - 2v_{F}^{2})\\
\Gamma^{\nu}\slashed{a}\slashed{b}\Gamma_{\nu} & = 4a_{\mu}b_{\nu}M^{\mu\nu} - \slashed{a}
\slashed{b}(3 - 2v_{F}^{2})\\
\Gamma^{\nu}\slashed{a}\slashed{b}\slashed{c}\Gamma_{\nu} & = - 2\slashed{c}\slashed{b}
\slashed{a} + \slashed{a}\slashed{b}\slashed{c}(3 - 2v_{F}^{2}),
\end{split}
\end{equation}
where $\slashed{a} = \Gamma^{\mu}a_{\mu}$.

\section{Form factors in anisotropic QED$_{2 + 1}$ at one-loop}\label{section 3}

\begin{figure}[!htb]
\includegraphics[scale=0.5]{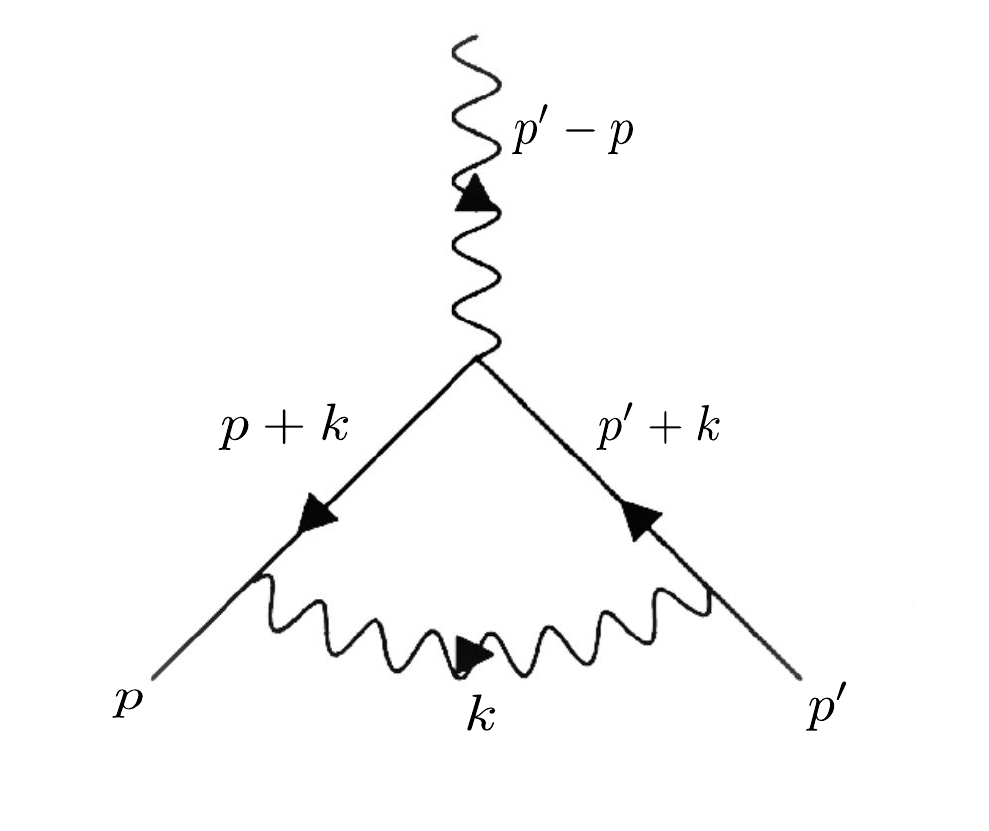}
\caption{Feynman diagram for the one-loop vertex correction.}
\label{FIG.2}
\end{figure}

In this section, we write down the expressions for the form factors in Proca quantum electrodynamics at the one-loop level.
It can be derived from the Feynman diagram presented in FIG. \ref{FIG.2}.
\\
Using the Feynman gauge for the photon's propagator, we find the scattering amplitude to be
\begin{equation}\label{eq. 4.1}
\begin{split}
ie\bar{\Gamma}_{\text{1 loop}}^{\mu}(p', p) & = \int_{reg}\frac{d^{3}k}{(2\pi)^{3}}
\frac{-ig^{\nu\lambda}}{k^{2} - m_{\gamma}^{2}}\times ie\Gamma_{\nu}\times\frac{i}{\slashed{p'} + 
\slashed{k} - m}\\
 & \times ie\Gamma^{\mu}\times\frac{i}{\slashed{p} + \slashed{k} - m}\times ie\Gamma_{\lambda}\\
  & = e^{3}\int_{reg}\frac{d^{3}k}{(2\pi)^{3}}\frac{\mathcal{N}^{\mu}}{\mathcal{D}},
\end{split}
\end{equation}
where
\begin{equation}\label{eq. 4.2}
\mathcal{N}^{\mu} = \Gamma^{\nu}(\slashed{k} + \slashed{p'} + m)\Gamma^{\mu}(\slashed{k} + 
\slashed{p} + m)\Gamma_{\nu},
\end{equation}
and
\begin{equation}\label{eq. 4.3}
\mathcal{D} = [k^{2} - m_{\gamma}^{2}]\times[(p + k)^{2} - m^{2}]\times[(p' + k)^{2} - m^{2}].
\end{equation}
After carrying out a long calculation (For detailed analysis, we refer to Appendix \ref{AppA}), we find that the gyromagnetic ratio is given by 
\begin{equation}\label{eq. 4.25}
\begin{split}
g & = - \frac{v_{F}^{2}}{m}\alpha\int_{0}^{1}\frac{w(1 - w)}{[w^{2} + (m_{\gamma}/m)^{2}
(1 - w)]^{3/2}} \ dw \\
 & \Bigg[\frac{2[2(1 - w) + v_{F}^{2}w]}{[v_{F}^{2} + (1 - w)(1 - v_{F}^{2})]^{2}}
 - \frac{4}{v_{F}^{2} + (1 - w)(1 - v_{F}^{2})}\Bigg]\\
 & = \frac{2v_{F}^{4}}{m}\alpha\int_{0}^{1} \frac{w^{2}(1 - w)}{[w^{2} + (m_{\gamma}/m)^{2}
(1 - w)]^{3/2}} \ dw\\
 & \times \frac{1}{[v_{F}^{2} + (1 - w)(1 - v_{F}^{2})]^{2}}.
\end{split}
\end{equation}
Numerically evaluating the above integral, we obtain the FIG.\ref{FIG.3} in which we have  plotted the gyromagnetic ratio $g$ as functions of Fermi velocity $v_{F}$ for three different values of mass $m_{\gamma}$ of gauge boson.
%
%
\\
\\
Next, we consider the momentum  of the center of mass of the incoming electrons to be zero. Then, the Pauli form factor can be expressed as
\begin{equation}\label{eq. 4.27}
\begin{split}
F_{2}(Q^{2}) & =  \frac{2v_{F}^{4}}{m}\alpha\int_{0}^{1}dw \frac{w^{2}(1 - w)}{[v_{F}^{2} + (1 - w)(1 - v_{F}^{2})]^{2}}\mathcal{I}_{1},
\end{split}
\end{equation}
where
\begin{equation}
\begin{split}
\mathcal{I}_{1} & = - \frac{4}{\sqrt{A},
(B - 4A)}\\
A & = w^{2} + \frac{m_{\gamma}^{2}}{m^{2}}(1 - w) + \frac{Q^{2}}{m^{2}}\left(\frac{v_{F}^{2}(1 - v_{F}^{2})w^{2}(1 - w)}{4[v_{F}^{2} + (1 - w)(1 - v_{F}^{2})]}\right),\\
B & = \frac{Q^{2}}{m^{2}}\frac{w^{2}(1 - v_{F}^{2})(v_{F}^{2} + 1 - w)}{v_{F}^{2} + (1 - w)(1 - v_{F}^{2})}. 
\end{split}
\end{equation}
Plotting this function numerically yields the FIG.\ref{FIG.5} where we show the  $\frac{Q}{m}$ dependence of Pauli form factor for three different values of mass ratios $\frac{m_{\gamma}}{m}$.
\\
\\
From the FIG.\ref{FIG.5}, we can see the Pauli form factor is weak when the Fermi velocity of electrons is $1/300$. However, if we consider ultra-relativistic fermions ($v_{F}\rightarrow 1$), then we obtain FIG.\ref{FIG.4} and FIG.\ref{FIG.6}. We find that both the gyromagnetic ratio and Pauli form factor become significant. It is evident from the figures that the g-factor and form factor increase as the Fermi velocity increases, whereas the relationship is inverse with photon mass. The form factors start decreasing as photon mass increases. 
\newpage
\begin{widetext}
\begin{figure}[!hb]
    \centering
    \begin{minipage}{.5\textwidth}
        \centering
		\includegraphics[scale=0.5]{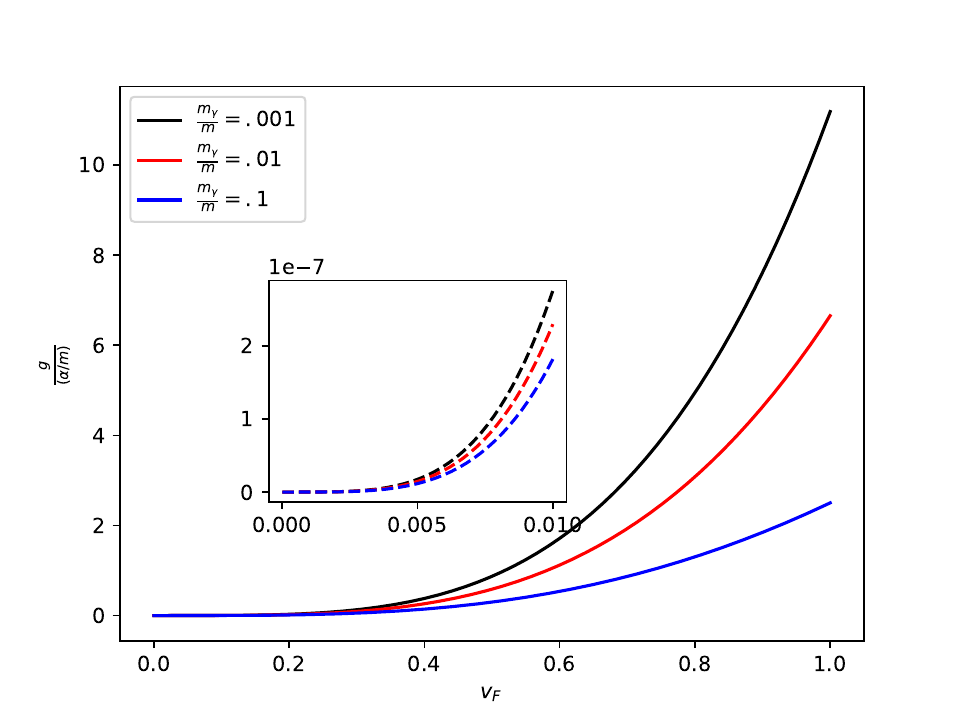}
		\caption{This figure shows how the gyromagnetic ratio changes with Fermi velocity $v_F$ for different photon masses.}
		\label{FIG.3}
    \end{minipage}%
    \hspace*{0.1in}
    \begin{minipage}{0.5\textwidth}
        \centering
		\includegraphics[scale=0.5]{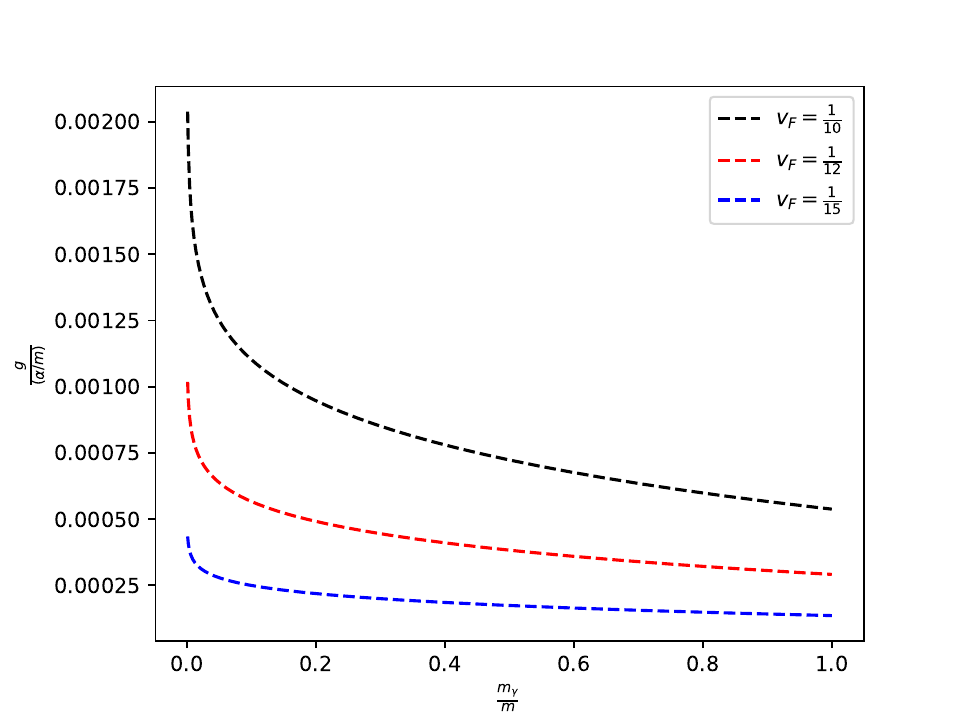}
		\caption{It describes the behaviour the gyromagnetic ratio with mass $m_{\gamma}$ of gauge boson for different $v_F$.}
		\label{FIG.4}
    \end{minipage}
\end{figure}
\begin{figure}
    \centering
    \begin{minipage}{.5\textwidth}
        \centering
		\includegraphics[scale=0.5]{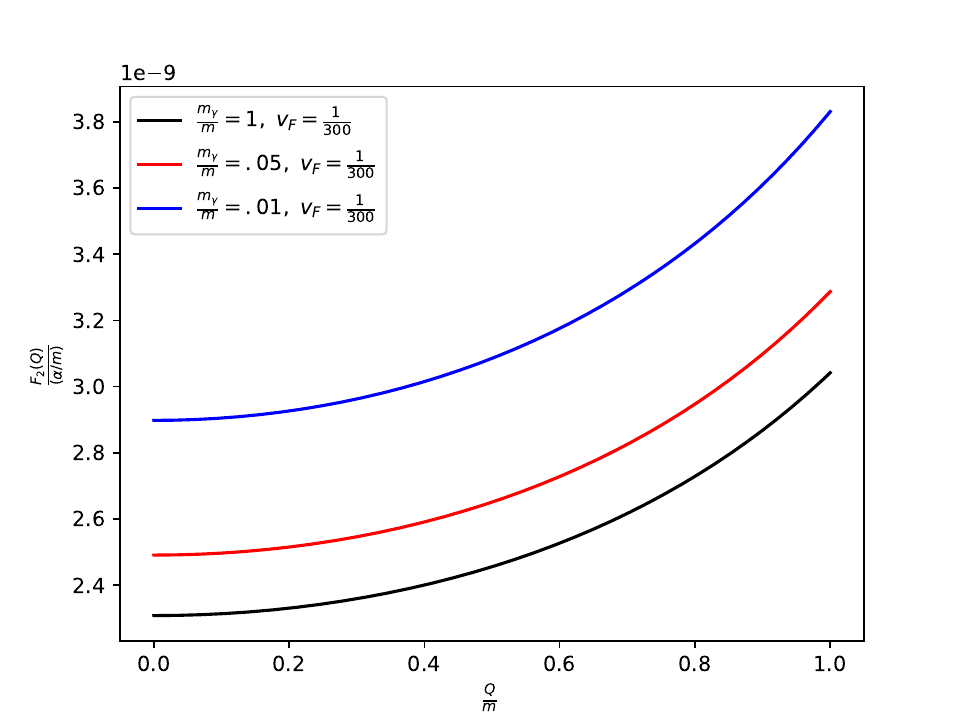}
		\caption{Dependence of Pauli form factor $F_2(Q)$ on the transfer momentum $Q$ for small $v_F$s.}
		\label{FIG.5}
    \end{minipage}%
    \hspace*{0.1in}
    \begin{minipage}{0.5\textwidth}
        \centering
		\includegraphics[scale=0.5]{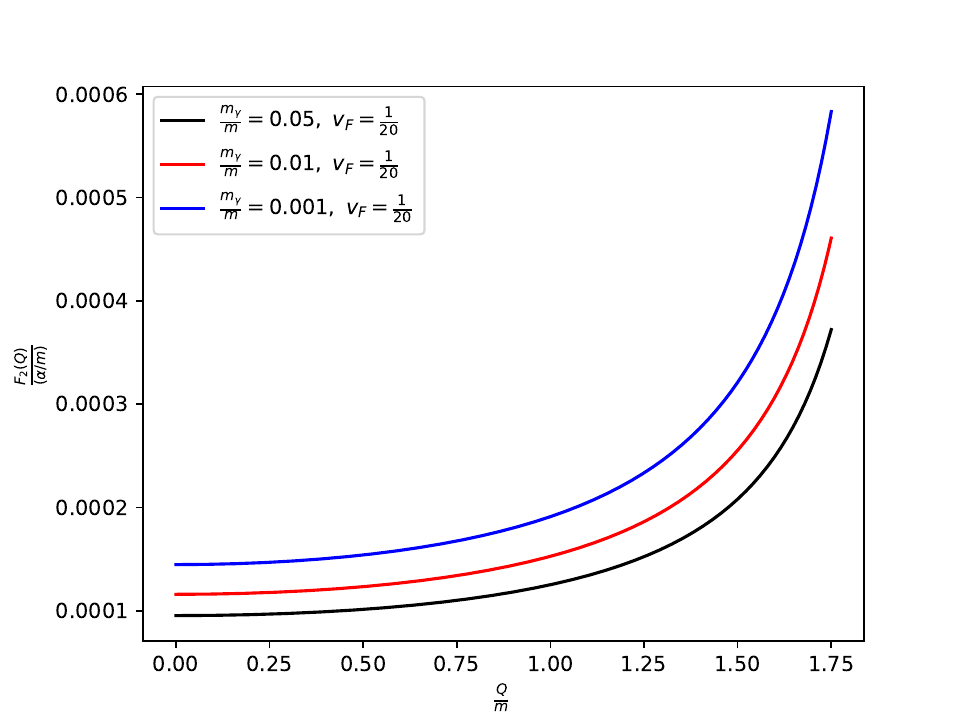}
		\caption{Dependence of Pauli form factor $F_2(Q)$ on the transfer momentum $Q$ in the ultra-relativistic limits.}
		\label{FIG.6}
    \end{minipage}
\end{figure}
\end{widetext}
\section{Discussion}
In summary, this paper discusses the spin gyromagnetic and Pauli form factors in the framework of anisotropic $(2+1)$D QED. In particular, we have determined these form factors for two-dimensional electronic systems with short-range interaction \cite{PhysRevB.102.125421, Tang2018, PhysRevB.109.165115, DasSarma2015} and investigated the effects of both the interaction and anisotropy on them, considering the Proca QED. We have found that the g-factor and Pauli form factor to be real and increase with Fermi velocity, whereas they decrease with the mass of the photon. 

\textbf{Data Availability Statement:} No Data associated in the manuscript.

\bibliographystyle{apsrev}
\bibliography{Draft.bbl}

\newpage

\appendix
\section{Detailed calculation for Form factors}
\label{AppA}
We have found that the interaction vertex is given by 
\begin{equation}\label{eq. 4.1}
\begin{split}
ie\bar{\Gamma}_{\text{1 loop}}^{\mu}(p', p) & = \int_{reg}\frac{d^{3}k}{(2\pi)^{3}}
\frac{-ig^{\nu\lambda}}{k^{2} - m_{\gamma}^{2}}\times ie\Gamma_{\nu}\times\frac{i}{\slashed{p'} + 
\slashed{k} - m}\\
 & \times ie\Gamma^{\mu}\times\frac{i}{\slashed{p} + \slashed{k} - m}\times ie\Gamma_{\lambda}\\
 & = e^{3}\int_{reg}\frac{d^{3}k}{(2\pi)^{3}}\frac{1}{k^{2} - m_{\gamma}^{2}}\times\Gamma^{\nu}
 \times\frac{\slashed{p'} + \slashed{k} + m}{(p' + k)^{2} - m^{2}}\\
 & \times\Gamma^{\mu}\frac{\slashed{p} + \slashed{k} + m}{(p + k)^{2} - m^{2}}\times\Gamma_{\nu}\\
 & = e^{3}\int_{reg}\frac{d^{3}k}{(2\pi)^{3}}\frac{\mathcal{N}^{\mu}}{\mathcal{D}},
\end{split}
\end{equation}
where the numerator and denominator are as follows:
\begin{equation}\label{eq. 4.2}
\mathcal{N}^{\mu} = \Gamma^{\nu}(\slashed{k} + \slashed{p'} + m)\Gamma^{\mu}(\slashed{k} + 
\slashed{p} + m)\Gamma_{\nu},
\end{equation}
and
\begin{equation}\label{eq. 4.3}
\mathcal{D} = [k^{2} - m_{\gamma}^{2}]\times[(p + k)^{2} - m^{2}]\times[(p' + k)^{2} - m^{2}].
\end{equation}
Before proceeding to the calculation, we write down a few relations which will be useful later in this calculation. They are
\begin{equation}
\begin{split}
(p + k)^{2} & = (p_{0} + k_{0})^{2} - v_{F}^{2}(\vec{p} + \vec{k})^{2}\\
(p' + k)^{2} & = (p'_{0} + k_{0})^{2} - v_{F}^{2}(\vec{p'} + \vec{k})^{2}.
\end{split}
\end{equation}
Using the Feynman trick, we now rewrite the denominator as
\begin{equation}\label{eq. 4.4}
\begin{split}
\frac{1}{\mathcal{D}} & = 2\int_{0}^{1}\int_{0}^{1}\int_{0}^{1}dx \ dy \ dz \ 
\delta(x + y + z - 1)\\
\times & \frac{1}{\Big[x((p + k)^{2} - m^{2}) + y((p' + k)^{2} - m^{2}) + z(k^{2} - m_{\gamma}^{2})\Big]^{3}}. 
\end{split}
\end{equation}
Next, we simplify the expression for denominator of the integrand in (\ref{eq. 4.4}). It leads to
\begin{equation}\label{eq. 4.5}
\begin{split}
[\ldots] & = x \times ((p + k)^{2} - m^{2}) + y \times ((p' + k)^{2} - m^{2})\\
 & + z \times (k^{2} - m_{\gamma}^{2})\\
 & = x[(p_{0} + k_{0})^{2} - v_{F}^{2}(\vec{p} + \vec{k})^{2} - m^{2}]\\
 & + y[(p'_{0} + k_{0})^{2} - v_{F}^{2}(\vec{p'} + \vec{k})^{2} - m^{2}] + z (k_{0}^{2} - 
 \vec{k}^{2} - m_{\gamma}^{2})\\
 & = k_{0}^{2}(x + y + z = 1) + x(p_{0}^{2} + 2p_{0}k_{0}) + y(p_{0}^{'2} + 2p'_{0}k_{0})\\
 & - v_{F}^{2}[x(\vec{p} + \vec{k})^{2} + y(\vec{p'} + \vec{k})^{2}] - z\vec{k}^{2} - m^{2}(1 - z)
 \\
 & = k_{0}^{2} + xp^{2} + yp^{'2} - \vec{k}^{2}(xv_{F}^{2} + yv_{F}^{2} + z) - zm_{\gamma}^{2}\\
 & + 2x(p_{0}k_{0} - v_{F}^{2}\vec{p}.\vec{k}) + 2y(p'_{0}k_{0} - v_{F}^{2}\vec{p'}.\vec{k})
 - m^{2}(1 - z)\\
 & = (k_{0} + xp_{0} + yp'_{0})^{2} - (xp_{0} + yp'_{0})^{2} + xp^{2} + yp^{'2}\\
 & - (v_{F}^{2} + z(1 - v_{F}^{2}))\Big\{\vec{k} + \frac{v_{F}^{2}}{v_{F}^{2} + z(1 - v_{F}^{2})}
 (x\vec{p} + y\vec{p'})\Big\}^{2}\\
 & + \frac{v_{F}^{4}}{v_{F}^{2} + z(1 - v_{F}^{2})}(x\vec{p} + y\vec{p'})^{2} - m^{2}(1 - z)
 - zm_{\gamma}^{2}.
\end{split}
\end{equation}
 Defining $q = p' - p$, we rearrange a few of the terms that are present in the above equation (\ref{eq. 4.5}) as following:
\begin{equation}\label{eq. 4.6}
\begin{split}
(xp_{0} + yp'_{0})^{2} & = x^{2}p_{0}^{2} + y^{2}p_{0}^{'2} - xy[(p'_{0} - p_{0})^{2} - 
p_{0}^{2} - p_{0}^{'2}]\\
 & = x(x + y)p_{0}^{2} + y(x + y)p_{0}^{'2} - xyq_{0}^{2}\\
(x\vec{p} + y\vec{p'})^{2} & = x(x + y)\vec{p}^{2} + y(x + y)\vec{p'}^{2} - xy\vec{q}^{2}. 
\end{split}
\end{equation}
As a result, (\ref{eq. 4.5}) further reduces to 
\begin{equation}\label{eq. 4.7}
\begin{split}
[\ldots] & = (k_{0} + xp_{0} + yp'_{0})^{2} + xzp_{0}^{2} + yzp_{0}^{'2} + xyq_{0}^{2}
 - zm_{\gamma}^{2}\\
 & - (v_{F}^{2} + z(1 - v_{F}^{2}))\Big\{\vec{k} + \frac{v_{F}^{2}}{v_{F}^{2} + z(1 - v_{F}^{2})}
 (x\vec{p} + y\vec{p'})\Big\}^{2}\\
 & - \frac{v_{F}^{2}}{v_{F}^{2} + z(1 - v_{F}^{2})}[xz\vec{p}^{2} + yz\vec{p'}^{2} + v_{F}^{2}xy
 \vec{q}^{2}] - m^{2}(1 - z)\\
 & = (k_{0} + xp_{0} + yp'_{0})^{2} + xyq_{0}^{2} - \frac{v_{F}^{4}xy\vec{q}^{2}}{v_{F}^{2} + 
 z(1 - v_{F}^{2})} - zm_{\gamma}^{2}\\
 & - (v_{F}^{2} + z(1 - v_{F}^{2}))\Big\{\vec{k} + \frac{v_{F}^{2}}{v_{F}^{2} + z(1 - v_{F}^{2})}
 (x\vec{p} + y\vec{p'})\Big\}^{2}\\
 & - \frac{v_{F}^{2}(1 - v_{F}^{2})(1 - z)}{v_{F}^{2} + z(1 - v_{F}^{2})}(xz\vec{p}^{2} + 
 yz\vec{p'}^{2}) - m^{2}(1 - z)^{2},
\end{split}
\end{equation}
where,  we have used fermion on-shell dispersions in the last line, i.e., $p_{0}^{2} - v_{F}^{2}\vec{p}^{2}
= m^{2}, \ p_{0}^{'2} - v_{F}^{2}\vec{p'}^{2} = m^{2}$.
We now turn to the expression for numerator. We find that
\begin{equation}\label{eq. 4.8}
\begin{split}
\mathcal{N}^{\mu} & = \Gamma^{\nu}(\slashed{k} + \slashed{p'} + m)\Gamma^{\mu}(\slashed{k} + 
\slashed{p} + m)\Gamma_{\nu}\\
 & = \frac{\partial}{\partial a_{\mu}}[\Gamma^{\nu}(\slashed{k} + \slashed{p'} + m)\slashed{a}(\slashed{k} + \slashed{p} + m)\Gamma_{\nu}]\\
 & = \frac{\partial}{\partial a_{\mu}}\Big[\Gamma^{\nu}(\slashed{k} + \slashed{p'})\slashed{a}
 (\slashed{k} + \slashed{p})\Gamma_{\nu} + m^{2}\Gamma^{\nu}\slashed{a}\Gamma_{\nu}\\
 & + m\Gamma^{\nu}\slashed{a}(\slashed{k} + \slashed{p})\Gamma_{\nu} + m\Gamma^{\nu}(\slashed{k} 
 + \slashed{p'})\slashed{a}\Gamma_{\nu}\Big]\\
 & = \frac{\partial}{\partial a_{\mu}}\Big[ - 2(\slashed{k} + \slashed{p})\slashed{a}(\slashed{k} 
 + \slashed{p'}) + (\slashed{k} + \slashed{p'})\slashed{a}(\slashed{k} + \slashed{p})(3 - 
 2v_{F}^{2})\\
 & + m^{2}\slashed{a}(1 - 2v_{F}^{2}) + 4ma_{\rho}(k + p)_{\sigma}M^{\rho\sigma} - m\slashed{a}
 (\slashed{k} + \slashed{p})(3 - 2v_{F}^{2})\\
 & + 4m(k + p')_{\rho}a_{\sigma}M^{\rho\sigma} - m(\slashed{k} + \slashed{p'})\slashed{a}(3 - 
 2v_{F}^{2})\Big]\\
 & = \Big[ - 2(\slashed{k} + \slashed{p})\Gamma^{\mu}(\slashed{k} 
 + \slashed{p'}) + (\slashed{k} + \slashed{p'})\Gamma^{\mu}(\slashed{k} + \slashed{p})(3 - 
 2v_{F}^{2})\\
 & + m^{2}\Gamma^{\mu}(1 - 2v_{F}^{2}) + 4m(2k + p + p')_{\sigma}M^{\mu\sigma}\\
 & - m\Gamma^{\mu}(\slashed{k} + \slashed{p})(3 - 2v_{F}^{2}) - m(\slashed{k} + \slashed{p'})
 \Gamma^{\mu}(3 - 2v_{F}^{2})\Big]\\
 & = \Big[ - 2(\slashed{k} + \slashed{p})\Gamma^{\mu}(\slashed{k} 
 + \slashed{p'}) - 2m^{2}\Gamma^{\mu}\\
  &+ (\slashed{k} + \slashed{p'} - m)\Gamma^{\mu}(\slashed{k} + \slashed{p} - m)(3 - 2v_{F}^{2})\\
 & + 4m(2k + p + p')_{\sigma}M^{\mu\sigma}\Big].
\end{split}
\end{equation}
Next, we shift the loop momentum from $k\rightarrow l$ such that
\begin{equation}\label{eq. 4.9}
\begin{split}
l_{0} & = k_{0} + xp_{0} + yp'_{0},\\
\vec{l} & = \vec{k} + \frac{v_{F}^{2}}{v_{F}^{2} + z(1 - v_{F}^{2})}(x\vec{p} + y\vec{p'}). 
\end{split}
\end{equation}
Together, they can be expressed as $l = k + s$, where $s^{\mu}$ is the shift vector. Inserting the $\bar{\Gamma}^{\mu}$ in-between $\bar{u}(p')$ and $u(p)$, we arrive at 
\begin{equation}\label{eq. 4.10}
\begin{split}
\mathcal{N}^{\mu} & = \Big[ - 2(\slashed{k} + \slashed{p})\Gamma^{\mu}(\slashed{k} 
 + \slashed{p'}) - 2m^{2}\Gamma^{\mu}\\
  & + \slashed{k}\Gamma^{\mu}\slashed{k}(3 - 2v_{F}^{2}) + m(2k + p + p')_{\sigma}M^{\mu\sigma}
  \Big]\\
  & = \Big[ - 2(\slashed{l} + \slashed{p} - \slashed{s})\Gamma^{\mu}(\slashed{l} + \slashed{p'}
  - \slashed{s}) - 2m^{2}\Gamma^{\mu}\\
  & + (\slashed{l} - \slashed{s})\Gamma^{\mu}(\slashed{l} - \slashed{s})(3 - 2v_{F}^{2})\\
  & + 4m(2l - 2s + p + p')_{\sigma}M^{\mu\sigma}\Big].
\end{split}
\end{equation} 
After a little algebra, it can be shown that the denominator of the integrand becomes quadratic in $l$. 
\begin{equation}\label{eq. 4.11}
\begin{split}
[\ldots] & = l_{0}^{2} + xyq_{0}^{2} - \frac{v_{F}^{4}xy\vec{q}^{2}}{v_{F}^{2} + 
 z(1 - v_{F}^{2})} - (v_{F}^{2} + z(1 - v_{F}^{2}))\vec{l}^{2}\\
 & - \frac{v_{F}^{2}(1 - v_{F}^{2})(1 - z)}{v_{F}^{2} + z(1 - v_{F}^{2})}(xz\vec{p}^{2} + 
 yz\vec{p'}^{2}) - m^{2}(1 - z)^{2} - zm_{\gamma}^{2}.
\end{split}
\end{equation}
The linear terms of $\mathcal{N}^{\mu}$ in $l$ (new loop momentum) do not contribute to the integral. As a result, the numerator takes the form
\begin{equation}\label{eq. 4.12}
\begin{split}
\mathcal{N}^{\mu} & = - 2m^{2}\Gamma^{\mu} + \slashed{l}\Gamma^{\mu}\slashed{l}(1 - 2v_{F}^{2})
 + \slashed{s}\Gamma^{\mu}\slashed{s}(3 - 2v_{F}^{2})\\
 & - 2(\slashed{p} - \slashed{s})\Gamma^{\mu}(\slashed{p'} - \slashed{s}) + 4m(p + p' - 2s)
 _{\sigma}M^{\mu\sigma}.
\end{split}
\end{equation}
In the next step, we utilize the following relations
\begin{equation}\label{eq. 4.13}
\begin{split}
p_{0} - s_{0} & = p_{0} - (xp_{0} + yp'_{0}) = zp'_{0} - (1 - x)q_{0}\\
p'_{0} - s_{0} & = p'_{0} - (xp_{0} + yp'_{0}) = zp_{0} + (1 - y)q_{0}\\
\vec{p} - \vec{s} & = \vec{p} - \frac{v_{F}^{2}}{v_{F}^{2} + z(1 - v_{F}^{2})}(x\vec{p} 
 + y\vec{p'})\\
 & = \frac{\vec{p'}z - \vec{q}[v_{F}^{2}(1 - x) + z(1 - v_{F}^{2})]}{v_{F}^{2} + z(1 - v_{F}^{2})}\\
\vec{p'} - \vec{s} & = \vec{p'} - \frac{v_{F}^{2}}{v_{F}^{2} + z(1 - v_{F}^{2})}(x\vec{p} 
 + y\vec{p'})\\
 & = \frac{\vec{p}z + \vec{q}[z + v_{F}^{2}x]}{v_{F}^{2} + z(1 - v_{F}^{2})}, 
\end{split}
\end{equation}
which yields
\begin{equation}\label{eq. 4.14}
\begin{split}
p_{0} + p'_{0} - 2s_{0} & = z(p_{0} + p'_{0}) + (x - y)q_{0}\\
\vec{p} + \vec{p'} - 2\vec{s} & = \frac{z(\vec{p} + \vec{p'}) + \vec{q}v_{F}^{2}(x - y)}{v_{F}^{2} 
 + z(1 - v_{F}^{2})}.
\end{split}
\end{equation}
Due to the $x\leftrightarrow y$ exchange symmetry of the Feynman parameters in the integral, we may further simplify
the above equations as
\begin{equation}\label{eq. 4.15}
p_{0} + p'_{0} - 2s_{0} \simeq z(p_{0} + p'_{0}), \ \vec{p} + \vec{p'} - 2\vec{s} \simeq 
\frac{z(\vec{p} + \vec{p'})}{v_{F}^{2} + z(1 - v_{F}^{2})}.
\end{equation}
We also note that
\begin{equation}\label{eq. 4.16}
\begin{split}
\slashed{l}\Gamma^{\mu}\slashed{l} & = l_{\lambda}\lambda_{\nu}\Gamma^{\nu}\Gamma^{\mu}
\Gamma^{\lambda}\simeq \frac{g_{\lambda\nu}l^{2}}{3}\Gamma^{\nu}\Gamma^{\mu}\Gamma^{\lambda}\\
 & = \frac{l^{2}}{3}\Gamma^{\nu}\Gamma^{\mu}\Gamma_{\nu} = \frac{l^{2}}{3}(2M_{ \ \nu}^{\mu}
 \Gamma^{\nu} - \Gamma^{\mu}(1 + 2v_{F}^{2})),
\end{split}
\end{equation}
and
\begin{equation}\label{eq. 4.16}
\begin{split}
s_{0} & = xp_{0} + yp'_{0} = (1 - z)p_{0} + yq_{0} = (1 - z)p'_{0} - q_{0}x\\
\vec{s} & = \frac{v_{F}^{2}(x\vec{p} + y\vec{p'})}{v_{F}^{2} + z(1 - v_{F}^{2})}
 = \frac{v_{F}^{2}((1 - z)\vec{p} + y\vec{q})}{v_{F}^{2} + z(1 - v_{F}^{2})}\\
 & = \frac{v_{F}^{2}((1 - z)\vec{p'} - \vec{q}x)}{v_{F}^{2} + z(1 - v_{F}^{2})}
\end{split}
\end{equation}
As a result, the first two terms in (\ref{eq. 4.12}) do not contribute to the $F_{2}$ (Pauli form factor). Furthermore, the numerator only depends on $p, p'$ and $q$. In the following step, we integrate over loop momentum which leads to
\begin{equation}\label{eq. 4.17}
\int\frac{d^{3}l}{(2\pi)^{3}}\frac{1}{[l_{0}^{2} - (v_{F}^{2} + z(1 - v_{F}^{2}))\vec{l}^{2} - 
\Delta]^{3}},
\end{equation}
where $\Delta$ is given by
\begin{equation}\label{eq. 4.18}
\begin{split}
\Delta & = - xyq_{0}^{2} + \frac{v_{F}^{4}xy\vec{q}^{2}}{v_{F}^{2} + z(1 - v_{F}^{2})} + 
\frac{v_{F}^{2}(1 - v_{F}^{2})(1 - z)}{v_{F}^{2} + z(1 - v_{F}^{2})}\\
 & \times (xz\vec{p}^{2} + yz\vec{p'}^{2}) + m^{2}(1 - z)^{2} + zm_{\gamma}^{2}.
 \end{split}
\end{equation}
The integral in (\ref{eq. 4.17})  evaluates to
\begin{equation}\label{eq. 4.19}
\begin{split}
2 & \int\frac{d^{2}l}{(2\pi)^{2}}\int_{0}^{\infty}\frac{dl_{0}}{2\pi}\frac{1}{[l_{0}^{2} - 
(v_{F}^{2} + z(1 - v_{F}^{2}))\vec{l}^{2} - \Delta]^{3}}\\
&= - \frac{3i}{8}  \int_{0}^{\infty}\frac{dl}{(2\pi)}\frac{l}{[l^{2}(v_{F}^{2} + z(1 - v_{F}^{2})) 
+ \Delta]^{5/2}},\\
&= - \frac{i}{16\pi}  \frac{1}{[v_{F}^{2} + z(1 - v_{F}^{2})]\Delta^{3/2}}.
\end{split}
\end{equation}
%
Now we consider the component $\bar{\mathcal{N}}^{i}$ that is defined as follows (note that here temporal components of 4-momenta are ignored) 
\begin{equation}\label{eq. 4.20}
\begin{split}
\bar{\mathcal{N}}^{i} & = \slashed{s}\Gamma^{i}\slashed{s}(3 - 2v_{F}^{2}) - 2(\slashed{p} - 
\slashed{s})\Gamma^{i}(\slashed{p'} - \slashed{s})\\
 &  + 4m(p + p' - 2s)_{j}M^{ij}\\
 & = s_{j}s_{k}\Gamma^{j}\Gamma^{i}\Gamma^{k}(3 - 2v_{F}^{2}) - \frac{2}{[v_{F}^{2} + z(1 - v_{F}^{2})]^{2}}\\
 & \times [mz - q_{j}\Gamma^{j}[v_{F}^{2}y + z]]\Gamma^{i}[mz + q_{k}\Gamma^{k}[z + v_{F}^{2}x]]\\
 & + \frac{4mv_{F}^{2}[z(p + p')^{i}]}{v_{F}^{2} + z(1 - v_{F}^{2})}\\
 & = (3 - 2v_{F}^{2})s_{j}s_{k}(2M^{ij}\Gamma^{k} - \Gamma^{i}\Gamma^{j}\Gamma^{k}) - 
 \frac{2}{[v_{F}^{2} + z(1 - v_{F}^{2})]^{2}}\\
 & \times\Big[m^{2}z^{2}\Gamma^{i} - mz(v_{F}^{2}y + z)q_{j}\Gamma^{j}\Gamma^{i} + mz[z + v_{F}^{2} 
 x]]q_{k}\Gamma^{i}\Gamma^{k}\\
 & - q_{j}q_{k}[v_{F}^{2}y + z][z + v_{F}^{2}x]]\Gamma^{j}\Gamma^{i}\Gamma^{k}\Big]\\
 & + \frac{4mv_{F}^{2}[z(p + p')^{i}]}{v_{F}^{2} + z(1 - v_{F}^{2})}.  
\end{split}
\end{equation}
Using the algebra of $\Gamma$ matrices we arrive at the above expression (\ref{eq. 4.20}), where the first term gives rise to a term proportional to $\Gamma$ matrix. Therefore, it will not contribute to gyromagnetic ratio. Similar situation arises for the first and the last terms inside the second square brackets. As a result, we are left with the following terms of
$\bar{\mathcal{N}}^{i}$, which we define as $\mathcal{N}_{2}^{i}$, i.e.,  
\begin{equation}\label{eq. 4.21}
\begin{split}
\mathcal{N}_{2}^{i} & = - \frac{2}{[v_{F}^{2} + z(1 - v_{F}^{2})]^{2}}\Big[mz(v_{F}^{2}y + z)
v_{F}^{2}q_{j}(i\sigma^{ji})\\
 & + mz[z + v_{F}^{2}x]v_{F}^{2}q_{k}(i\sigma^{ki})\Big] - \frac{4mv_{F}^{2}z}{v_{F}^{2} + z(1 - v_{F}^{2})}(i\sigma^{ij})q_{j}.
\end{split}
\end{equation}
Therefore, the gyromagnetic ratio is given by 
\begin{equation}\label{eq. 4.22}
\begin{split}
g & = - m\frac{e^{2}}{4\pi}\int_{0}^{1}\int_{0}^{1}\int_{0}^{1}dx \ dy \ dz \ \delta(x + y + z - 1)\\
 & \times\frac{1}{[v_{F}^{2} + z(1 - v_{F}^{2})]\Delta^{3/2}}\Bigg[\Big[mzv_{F}^{2}(v_{F}^{2}y + z) + 
  mz[z + v_{F}^{2}x]v_{F}^{2}\Big]\\
 & \times\frac{2}{[v_{F}^{2} + z(1 - v_{F}^{2})]^{2}} - \frac{4mv_{F}^{2}z}{v_{F}^{2} + z(1 - v_{F}^{2})}\Big],
\end{split}
\end{equation}
where $\Delta$ is evaluated at $p = p' = q = 0$, leading to the expression
\begin{equation}\label{eq. 4.23}
\Delta = m^{2}(1 - z)^{2} + zm_{\gamma}^{2}.
\end{equation}
Next, we perform change of variables, given by
\begin{equation}\label{eq. 4.24}
x = w\xi, \ y = w(1 - \xi), \ z = 1 - w.
\end{equation}
Finally, we obtain following expression for the gyromagnetic ratio
\begin{equation}\label{eq. 4.25}
\begin{split}
g & = - \frac{v_{F}^{2}}{m}\alpha\int_{0}^{1}\frac{w(1 - w)}{[w^{2} + (m_{\gamma}/m)^{2}
(1 - w)]^{3/2}} \ dw \\
 & \Bigg[\frac{2[2(1 - w) + v_{F}^{2}w]}{[v_{F}^{2} + (1 - w)(1 - v_{F}^{2})]^{2}}
 - \frac{4}{v_{F}^{2} + (1 - w)(1 - v_{F}^{2})}\Bigg]\\
 & = \frac{2v_{F}^{4}}{m}\alpha\int_{0}^{1} \frac{w^{2}(1 - w)}{[w^{2} + (m_{\gamma}/m)^{2}
(1 - w)]^{3/2}} \ dw\\
 & \times \frac{1}{[v_{F}^{2} + (1 - w)(1 - v_{F}^{2})]^{2}}.
\end{split}
\end{equation}
After numerically evaluating the above integral, we get the FIG.\ref{FIG.3}, which shows how the gyromagnetic ratio $g$ behaves as function of Fermi velocity $v_{F}$ for three different values of mass $m_{\gamma}$ of gauge boson.
%
%
It is important to note that $\Delta$ generally depends on $q_{0}, \vec{q}$ and $\vec{p},  \vec{p'}$ due to the anisotropy. Next, we consider the on-shell condition of outgoing photon, which leads to the relation $q_{0}^{2} = \vec{q}^{2} = Q^{2}$. In addition, we also assume the center of mass momentum of the incoming electrons to be zero. Then, the function $\Delta$ can be expressed as
\begin{equation}\label{eq. 4.26}
\begin{split}
\Delta & = m^{2}(1 - z)^{2} + m_{\gamma}^{2}z + Q^{2}\Big[ - xy + \frac{v_{F}^{4}xy}{v_{F}^{2}
 + z(1 - v_{F}^{2})}\\
 & + \frac{v_{F}^{2}(1 - v_{F}^{2})(1 - z)^{2}z}{4[v_{F}^{2} + z(1 - v_{F}^{2})]}\Big]\\
 & = m^{2}(1 - z)^{2} + m_{\gamma}^{2}z + Q^{2}\Big[\frac{v_{F}^{2}(1 - v_{F}^{2})(1 - z)^{2}z}
 {4[v_{F}^{2} + z(1 - v_{F}^{2})]}\\
 & - \frac{xy(1 - v_{F}^{2})(v_{F}^{2} + z)}{v_{F}^{2} + z(1 - v_{F}^{2})}\Big]\\
 & = m^{2}w^{2} + m_{\gamma}^{2}(1 - w) + Q^{2}\Big[\frac{v_{F}^{2}(1 - v_{F}^{2})w^{2}(1 - w)}
 {4[v_{F}^{2} + (1 - w)(1 - v_{F}^{2})]}\\
 & - \frac{w^{2}\xi(1 - \xi)(1 - v_{F}^{2})(v_{F}^{2} + 1 - w)}{v_{F}^{2} + (1 - w)(1 - v_{F}^{2})
 }\Big]\\
\bar{\Delta} & = \frac{\Delta}{m^{2}} = w^{2} + \frac{m_{\gamma}^{2}}{m^{2}}(1 - w)
 + \frac{Q^{2}}{m^{2}}\Big[\frac{v_{F}^{2}(1 - v_{F}^{2})w^{2}(1 - w)}{4[v_{F}^{2} 
 + (1 - w)(1 - v_{F}^{2})]}\\
 & - \frac{w^{2}\xi(1 - \xi)(1 - v_{F}^{2})(v_{F}^{2} + 1 - w)}{v_{F}^{2} + (1 - w)(1 - v_{F}^{2})
 }\Big]. 
\end{split}
\end{equation} 
Finally, we find that the Pauli form is given by 
\begin{equation}\label{eq. 4.27}
\begin{split}
F_{2}(Q^{2}) & = \frac{2v_{F}^{4}}{m}\alpha\int_{0}^{1}d\xi\int_{0}^{1}dw \frac{w^{2}(1 - w)}{\bar{\Delta}(\xi, w)^{3/2}}\frac{1}{[v_{F}^{2} + (1 - w)(1 - v_{F}^{2})]^{2}}\\
 & = \frac{2v_{F}^{4}}{m}\alpha\int_{0}^{1}dw \frac{w^{2}(1 - w)}{[v_{F}^{2} + (1 - w)(1 - v_{F}^{2})]^{2}}\mathcal{I}_{1}\\
\mathcal{I}_{1} & = \int_{0}^{1}d\xi \ \frac{1}{[\bar{\Delta}(w,\xi)]^{3/2}} = - \frac{4}{\sqrt{A}
(B - 4A)}\\
A & = w^{2} + \frac{m_{\gamma}^{2}}{m^{2}}(1 - w) + \frac{Q^{2}}{m^{2}}\left(\frac{v_{F}^{2}(1 - v_{F}^{2})w^{2}(1 - w)}{4[v_{F}^{2} + (1 - w)(1 - v_{F}^{2})]}\right)\\
B & = \frac{Q^{2}}{m^{2}}\frac{w^{2}(1 - v_{F}^{2})(v_{F}^{2} + 1 - w)}{v_{F}^{2} + (1 - w)(1 - v_{F}^{2})}. 
\end{split}
\end{equation}

\end{document}